\documentclass[aps,pra,onecolumn,groupedaddress,showpacs]{revtex4}

\usepackage{amssymb,amsmath}
\usepackage{graphicx}

\newcommand{\om}{\omega}

\newcommand{\pa}{\partial}

\begin{document}

\title{Two-component nonlinear wave of the Born-Infeld equation}

\author{G. T. Adamashvili}
\affiliation{Technical University of Georgia, Kostava str.77, Tbilisi, 0179, Georgia.\\ email: $guram_{-}adamashvili@ymail.com.$ }

\begin{abstract} The generalized perturbative reduction method is used to find the two-component vector breather solution of the Born-Infeld equation
$ U_{tt} -C U_{zz} = - A U_{t}^{2} U_{zz} - \sigma U_{z}^{ 2} U_{tt} + B U_{z} U_{t} U_{zt} $. It is shown that the solution of the two-component nonlinear wave
oscillates with the sum and difference of frequencies and wave numbers.

\vskip+0.2cm
\emph{Keywords:} Generalized perturbative reduction method, Two-component nonlinear waves, Born-Infeld equation.
\end{abstract}


\maketitle

\section{Introduction}

The nonlinear solitary waves   plays a fundamental role in the study of nonlinear phenomena in completely different fields of physics and applied mathematics. The solitary wave behavior can mathematically be described in nonlinear partial differential equations. Some of the equations are the Boussinesq, Benjamin-Bona-Mahony, Maxwell-Bloch, Hirota and many others.
The nonlinear solitary waves  can be divided into two main types: single-component (scalar) and two-component (vector) solitary waves.
This is one of the most interesting topics for study in various fields of physics,various fields of physics such as optics, hydrodynamics, acoustics, plasma and others[1-5].

Among other nonlinear partial differential equations, there is a nonlinear modification of the Maxwell wave equation, which includes the so-called Born-Infeld nonlinearity. As a result, the Born-Infeld equation is obtained, which describes the properties of particles and has the form [1,6]
\begin{equation}\label{eq1}
\frac{\partial^{2} U}{\partial t^{2}}-C \frac{\partial^{2} U}{\partial z^{2}} =-  A (\frac{\partial U}{\partial t})^{2}   \frac{\partial^{2} U}{\partial z^{2}} - \sigma
(\frac{\partial U}{\partial z})^{2}   \frac{\partial^{2} U}{\partial t^{2}}     + B \frac{\partial U}{\partial t}\frac{\partial U}{\partial z} \frac{\partial^{2} U}{\partial t   \partial z},
\end{equation}
or in the dimensionless form
\begin{equation}\label{eqo}
 \frac{\partial^{2} U}{\partial t^{2}}-\frac{\partial^{2} U}{\partial z^{2}} =- (\frac{\partial U}{\partial t})^{2}   \frac{\partial^{2} U}{\partial z^{2}}-
(\frac{\partial U}{\partial z})^{2}   \frac{\partial^{2} U}{\partial t^{2}}     + 2 \frac{\partial U}{\partial t}\frac{\partial U}{\partial z} \frac{\partial^{2} U}{\partial t   \partial z},
\end{equation}
where $U(z,t)$ is a real function of space coordinate $z$ and time $t$ and represents the wave profile, while  $A,\;B,\;C$ and $\sigma $ are the real constants.
Eq.(2) has solutions in the form of the solitary waves $U=\Phi(x+t)$ and $U=\Phi(x-t)$, where  $\Phi$ is arbitrary function[1,6,7].

Sometimes 2+1 (two space coordinates and time) dimensional version of the Born-Infeld equation also have been investigated[8].

We shall consider a pulse with width $T$, the carrier frequency $\omega$ and wave number $k$, propagating along the positive $z$ axis.
We are interested in the case where the pulse duration is much longer than the inverse frequency of the carrier wave, i.e. $T>>1/\omega$.
Following the standard procedure of the slowly varying envelope approximation, we will transform the Born-Infeld equation \eqref{eq1} for real function $U$ into the slowly varying envelope functions, using the expansion [9,10]
\begin{equation}\label{eq2}
U(z,t)=\sum_{l=\pm1}\hat{u}_{l}(z,t) Z_l,
\end{equation}
where $Z_{l}= e^{il(k z -\omega t)}$ is the fast oscillating function, $\hat{u}_{l}$ are the slowly varying complex envelope functions, which satisfied inequalities
\begin{equation}\label{swa}
 \left|\frac{\partial \hat{u}_{l}}{\partial t}\right|\ll\omega
|\hat{u}_{l}|,\;\;\;\left|\frac{\partial \hat{u}_{l}}{\partial z
}\right|\ll k|\hat{u}_{l}|.
\end{equation}
For the reality of $U$, we set: $ \hat{u}_{+1}= \hat{u}^{*}_{-1}$.

The purpose of the present work is to consider the two-component vector breather solution of the Born-Infeld equation (1) using the generalized perturbative reduction method [3,4,11-16].

The rest of this paper is organized as follows: Section II is devoted to the linear part of the Born-Infeld equation for slowly varying complex envelope functions. In Section III, using the generalized perturbative reduction method, we will transform Eq.(1) to the coupled nonlinear  Schr\"odinger equations for auxiliary functions.
In Section IV, will be presented the solution of the two-component nonlinear pulse. Finally, in Section V, we will discuss the obtained results.

\vskip+0.5cm

\section{The linear part of the Born-Infeld equation}

The linear part of the Born-Infeld equation (1) is coincide with the linear Maxwell wave equation and is given by
\begin{equation}\label{lin}
\frac{\pa^{2} U}{\pa t^2} -C \frac{\pa^{2} U}{\pa z^{2}}=0.
\end{equation}

We consider a pulse whose duration satisfies  the condition $T >>\omega^{-1}$.  Substituting  Eq.(3) into (5) we obtain the dispersion relation
\begin{equation}\label{dis}
{\omega}^{2}= C k^{2}
\end{equation}
and the rest part of the linear equation (5) in the form
\begin{equation}\label{lin2}
\sum_{l=\pm1}Z_l [-2il\omega \frac{\partial \hat{u}_{l}}{\partial
t}- 2ilkC \frac{\partial \hat{u}_{l}}{\partial z}+\frac{\partial^{2} \hat{u}_{l}}{\partial t^{2}} -C
\frac{\partial^{2} \hat{u}_{l}}{\partial z^{2}})]=0.
\end{equation}

\vskip+0.5cm

\section{The generalized perturbative reduction method}

In order to consider the two-component vector breather solution of the Eq.(1), we use the generalized perturbative reduction method [3,4,11-16] which makes it possible to transform the Born-Infeld  equation for the functions $\hat{u}_{l}$ to the coupled nonlinear Schr\"odinger  equations for auxiliary functions $f_{l,n}^ {(\alpha)}$.
As a result, we obtain a two-component nonlinear pulse oscillating with the  difference and sum of the frequencies and wave numbers. In the frame of this method, the complex envelope function  $\hat{u}_{l}$ can be represented as
\begin{equation}\label{gprm}
\hat{u}_{l}(z,t)=\sum_{\alpha=1}^{\infty}\sum_{n=-\infty}^{+\infty}\varepsilon^\alpha
Y_{l,n} f_{l,n}^ {(\alpha)}(\zeta_{l,n},\tau),
\end{equation}
where $\varepsilon$ is a small parameter,
$$
Y_{l,n}=e^{in(Q_{l,n}z-\Omega_{l,n}
t)},\;\;\;\zeta_{l,n}=\varepsilon Q_{l,n}(z-v_{{g;}_{l,n}} t),
$$$$
\tau=\varepsilon^2 t,\;\;\;
v_{{g;}_{l,n}}=\frac{\partial \Omega_{l,n}}{\partial Q_{l,n}}.
$$

It is assumed that the quantities $\Omega_{l,n}$, $Q_{l,n}$ and $f_{l,n}^{(\alpha)}$ satisfies the inequalities for any $l$ and $n$:
\begin{equation}\label{rtyp}\nonumber\\
\omega\gg \Omega_{l,n},\;\;k\gg Q_{l,n},\;\;\;
\end{equation}
$$
\left|\frac{\partial
f_{l,n}^{(\alpha )}}{
\partial t}\right|\ll \Omega_{l,n} \left|f_{l,n}^{(\alpha)}\right|,\;\;\left|\frac{\partial
f_{l,n}^{(\alpha )}}{\partial \eta }\right|\ll Q_{l,n} \left|f_{l,n}^{(\alpha )}\right|.
$$

Substituting Eq.(8) into (7), for the linear part of the Born-Infeld  equation (5) we obtain
\begin{equation}\label{eqz}
\sum_{l=\pm1}\sum_{\alpha=1}^{\infty}\sum_{n=\pm 1}\varepsilon^\alpha Z_{l} Y_{l,n}[W_{l,n}
+\varepsilon J_{l,n} - \varepsilon^2 i l h_{l,n}  \frac{\partial }{\partial \tau}
-\varepsilon^{2} Q^{2} H_{l,n}\frac{\partial^{2} }{\partial \zeta^{2}}+O(\varepsilon^{3})]f_{l,n}^{(\alpha)}=0,
\end{equation}
where
\begin{equation}\label{cof}
W_{l,n}=- 2 n l\omega \Omega_{l,n}  + 2  n  l k Q_{l,n} C - \Omega^{2}_{l,n}+C  Q_{l,n}^{2},
$$
$$
J_{l,n}=2i Q_{l,n} [l \omega  v_{{g;}_{l,n}}   -l k C   + n \Omega_{l,n}  v_{{g;}_{l,n}}   -C n Q_{l,n}],
$$
$$
h_{l,n}=2(\omega + ln \Omega_{l,n}),
$$
$$
H_{l,n}=   C- v_{{g;}_{l,n}}^{2}.
\end{equation}

Equating to zero, the terms with the same powers of $\varepsilon$, from the Eq.(9) we obtain a series of equations. In the first order of $\varepsilon$, we have
a connection between of the parameters $\Omega_{l,n}$ and $Q_{l,n}$. When
\begin{equation}\label{fo1}
2 ( C k Q_{\pm1, \pm1} -\omega \Omega_{\pm1, \pm1}) - \Omega^{2}_{\pm1, \pm1} + C  Q_{\pm1, \pm1}^{2}=0,
\end{equation}
than $ f_{\pm1, \pm1}^{(1)}\neq0$  and when
\begin{equation}\label{fo2}
2 ( C k Q_{\pm1, \mp1} -\omega \Omega_{\pm1, \mp1}) + \Omega^{2}_{\pm1, \mp1} - C  Q_{\pm1, \mp1}^{2}=0,
\end{equation}
than $ f_{\pm1, \mp1}^{(1)}\neq0$.

From Eq.(10), in  the second order of $\varepsilon$, we obtain the equation
\begin{equation}\label{jo}\nonumber
J_{\pm1, \pm1}=J_{\pm1, \mp1}=0
\end{equation}
and the expression
\begin{equation}\label{v}
v_{{g;}_{l,n}}=C \frac{  k + l n Q_{l,n} }{ \omega   +l n  \Omega_{l,n} }.
\end{equation}

In the third order of $\varepsilon$,  the linear part of the Born-Infeld  equation (1), is given by
\begin{equation}\label{l}
\sum_{l=\pm1}\sum_{n=\pm 1}\varepsilon^{3} Z_{l} Y_{l,n}[ - i l h_{l,n}  \frac{\partial }{\partial \tau}- Q_{l,n}^{2} H_{l,n}\frac{\partial^{2} }{\partial \zeta^{2}}].
\end{equation}

Next we consider the nonlinear term of the  Born-Infeld  equation (1)
\begin{equation}\label{nonl}
-  A (\frac{\partial U}{\partial t})^{2}   \frac{\partial^{2} U}{\partial z^{2}} - \sigma
(\frac{\partial U}{\partial z})^{2}   \frac{\partial^{2} U}{\partial t^{2}}     + B \frac{\partial U}{\partial t}\frac{\partial U}{\partial z} \frac{\partial^{2} U}{\partial t   \partial z}.
\end{equation}

Substituting Eqs.(3) and (8) into Eq.(15), for the nonlinear part of the Born-Infeld  equation (1) we obtain
\begin{equation}\label{nonlz}
  \varepsilon^{3}\;Z_{+1}[(\tilde{q}_{+} | f_{+1,+1}^ {(1)}|^{2} + \tilde{r}_{+} | f_{+1,-1}^ {(1)}|^{2} ) f_{+1,+1}^ {(1)}Y_{+1,+1}
   + (\tilde{q}_{-} |f_{+1,-1}^ {(1)}|^{2} + \tilde{r}_{-} |f_{+1,+1} ^ {(1)}|^{2} ) Y_{+1,-1} f_{+1,-1}^ {(1)}]
 \end{equation}
and plus terms proportional to $ Z_{-1}$.
Here
$$
\tilde{q}_{\pm}=(A +\sigma -B )(\om \pm \Omega_{\pm } )^{2}  (k\pm  Q_{\pm} )^{2},
$$
$$
\tilde{r}_{\pm}=2[ A  (k\pm  Q_{\pm} )^{2}  (\om \mp\Omega_{\mp} )^{2}    +  \sigma  ( \omega \pm \Omega_{\pm }  )^{2}     (k \mp Q_{\mp}  )^{2}
-B     (  \om + \Omega_{+}  )   (  \om -\Omega_{-} ) (  k+ Q_{+}  )  (   k- Q_{-} )].
$$

In the third order of $\varepsilon$ for the Born-Infeld  equation (1), from Eqs.(14) and (16)  we obtain the system of nonlinear equations
\begin{equation}\label{2eq}
  i \frac{\partial f_{+1,+1}^{(1)}}{\partial \tau} + Q_{+}^{2} \frac{H_{+1,+1} }{h_{+1,+1}} \frac{\partial^2 f_{+1,+1}^{(1)}}{\partial \zeta_{+1,+1} ^2}+(\frac{\tilde{q}_{+}}{ h_{+1,+1}}   | f_{+1,+1}^ {(1)}|^{2} + \frac{\tilde{r}_{+}}{ h_{+1,+1}} | f_{+1,-1}^ {(1)}|^{2} ) f_{+1,+1}^ {(1)}=0,
$$$$
   i \frac{\partial f_{+1,-1 }^{(1)}}{\partial \tau} + Q_{-}^{2} \frac{H_{+1,-1} }{h_{+1,-1}} \frac{\partial^2 f_{+1,-1 }^{(1)}}{\partial \zeta_{+1,-1}^2}+(\frac{\tilde{q}_{-}}{h_{+1,-1}}  |f_{+1,-1}^ {(1)}|^{2} +\frac{\tilde{r}_{-}}{h_{+1,-1}} |f_{+1,+1} ^ {(1)}|^{2} )  f_{+1,-1}^ {(1)}=0.
 \end{equation}

\vskip+0.5cm

\section{The two-component vector breather of the Born-Infeld  equation}

Taking into account Eqs.(8) and (13), after transformation back to the space coordinate $z$ and time $t$, from the system of equations (17) we obtain the coupled nonlinear Schr\"odinger equations for the auxiliary functions $\Lambda_{\pm}=\varepsilon  f_{+1,\pm1}^{(1)}$ in the following form
\begin{equation}\label{pp2}
i (\frac{\partial \Lambda_{\pm}}{\partial t}+v_{\pm} \frac{\partial  \Lambda_{\pm}} {\partial z}) + p_{\pm} \frac{\partial^{2} \Lambda_{\pm} }{\partial z^{2}}
+q_{\pm}|\Lambda_{\pm}|^{2}\Lambda_{\pm} +r_{\pm} |\Lambda_{\mp}|^{2} \Lambda_{\pm}=0,
\end{equation}
where
\begin{equation}\label{OmQ}
p_{\pm}=\frac{ C- v_{\pm}^{2} }{2(\omega \pm \Omega_{\pm})},
$$
$$
 q_{\pm}=\frac{ \tilde{q}_{\pm}}{2(\omega \pm \Omega_{\pm})},
$$
$$
r_{\pm}=\frac{ \tilde{r}_{\pm}}{2(\omega \pm \Omega_{\pm})},
$$
$$
v_{\pm }= v_{g;_{+1,\pm 1}}=C \frac{  k \pm  Q_{\pm} }{ \omega   \pm   \Omega_{\pm } },
$$
$$
 \Omega_{+}=\Omega_{+1,+1}= \Omega_{-1,-1},\;\;\;\;\;\;\;\;\;\;\;\;\;\;\;\;\;\;\;\;\;\;\;\;\;  \Omega_{-}= \Omega_{+1,-1}= \Omega_{-1,+1},
$$
$$
 Q_{+}=Q_{+1,+1}= Q_{-1,-1},\;\;\;\;\;\;\;\;\;\;\;\;\;\;\;\;\;\;\;\;\;\;\;\;\;  Q_{-}= Q_{+1,-1}= Q_{-1,+1}.
\end{equation}

The solution of Eq.(18) is given by [3,11-14]
\begin{equation}\label{19}
\Lambda_{\pm }=\frac{A_{\pm }}{T}Sech(\frac{t-\frac{z}{V_{0}}}{T}) e^{i(k_{\pm } z - \omega_{\pm } t )},
\end{equation}
where $A_{\pm },\; k_{\pm }$ and $\omega_{\pm }$ are the real constants, $V_{0}$ is the velocity of the nonlinear wave. We assume that
$k_{\pm }<<Q_{\pm }$  and $\omega_{\pm }<<\Omega_{\pm }.$

Combining Eqs.(3), (8) and (20), we obtain the two-component vector breather solution of the Born-Infeld  equation (1) in the following form:
\begin{equation}\label{vb}
U(z,t)=\mathfrak{A} Sech(\frac{t-\frac{z}{V_{0}}}{T})\{  \cos[(k+Q_{+}+k_{+})z -(\omega +\Omega_{+}+\omega_{+}) t]
$$$$
+(\frac{p_{-}q_{+}-p_{+}r_{-}} {p_{+}q_{-}- p_{-}r_{+}})^{\frac{1}{2}} \cos[(k-Q_{-}+k_{-})z -(\omega -\Omega_{-}+\omega_{-})t]\},
\end{equation}
where $\mathfrak{A}$ is amplitude of the nonlinear pulse. The expressions for  the parameters $k_{\pm }$ and $\omega_{\pm }$  are given by
\begin{equation}\label{rt16}
k_{\pm }=\frac{V_{0}-v_{\pm}}{2p_{\pm}},\;\;\;\;\;\;\;\;\;\;\;\;\;\;\;\;\;\;\;\;\;
\omega_{+}=\frac{p_{+}}{p_{-}}\omega_{-}+\frac{V^{2}_{0}(p_{-}^{2}-p_{+}^{2})+v_{-}^{2}p_{+}^{2}-v_{+}^{2}p_{-}^{2}
}{4p_{+}p_{-}^{2}}.
\end{equation}

\vskip+0.5cm

\section{Conclusion}

We investigate the two-component vector breather solution of the Born-Infeld equation (1) in case when the slowly varying envelope approximation Eq.(4) is valid.
The nonlinear pulse with the width $T>>\Omega_{\pm }^{-1}>>\omega^{-1}$ is considered. Using the generalized perturbative reduction method Eq.(8), the Eq.(1) is  transformed to the coupled nonlinear Schr\"odinger equations (18) for the functions $\Lambda_{\pm 1}$. As a result, the two-component nonlinear pulse oscillating with the sum and difference  of the frequencies and wave numbers Eq.(21), is obtained. The dispersion relation and the connection  between parameters $\Omega_{\pm}$ and $Q_{\pm}$ are determined from Eqs.(6), (11) and (12). The parameters of the pulse from Eqs. (19) and (22) are determined.

We have to note that the two-component vector breathers can propagate also in the other physical systems\cite{Adamashvili:CPB:21, Adamashvili:OS:2019, Adamashvili:AcousPhy:17, Adamashvili:PhysRevE:12}.

\vskip+0.5cm

\end{document}